\documentclass[aps,pre,onecolumn,notitlepage,superscriptaddress,nogroupedaddress]{revtex4-2}

\usepackage{xcolor,graphicx}
\usepackage{amssymb,amsmath,graphicx}
\usepackage{mathtools}
\usepackage[utf8]{inputenc}
\usepackage{comment}
\usepackage{braket}
\usepackage[normalem]{ulem}
\usepackage{hyperref,url}

\usepackage{cleveref}
\usepackage{physics}

\newcommand{\deletetext}[1]{\iffalse{{\color{red}{#1}}}\fi}
\newcommand{\newtext}[1]{{{#1}}}

\graphicspath{{../}}

\begin{document}

\title{Convergence of the Integral Fluctuation Theorem estimator for nonequilibrium Markov systems}




\author{Francesco Coghi}
\email{francesco.coghi@su.se}
\affiliation{Nordita, KTH Royal Institute of Technology and Stockholm University, Hannes Alfvéns väg 12, SE-106 91 Stockholm, Sweden}

\author{Lorenzo Buffoni}
\email{lbuffoni@lx.it.pt}
\affiliation{Portuguese Quantum Institute, 1049-001 Lisbon, Portugal}

\author{Stefano Gherardini}
\email{stefano.gherardini@ino.cnr.it}
\affiliation{Istituto Nazionale di Ottica -- CNR, Area Science Park, Basovizza, I-34149 Trieste, Italy}
\affiliation{LENS, University of Florence, via Carrara 1, I-50019 Sesto Fiorentino, Italy}

\begin{abstract}
The Integral Fluctuation Theorem for entropy production (IFT) is among the few equalities that are known to be valid for physical systems arbitrarily driven far from equilibrium. Microscopically, it can be understood as an inherent symmetry for the fluctuating entropy production rate implying the second law of thermodynamics. Here, we examine an IFT statistical estimator based on regular sampling and discuss its limitations for nonequilibrium systems, when sampling rare events becomes pivotal. Furthermore, via a large deviation study, we discuss a method to carefully setup an experiment in the parameter region where the IFT estimator safely converges and also show how to improve the convergence region for Markov chains with finite correlation \deletetext{length} \newtext{time}. We corroborate our arguments with two illustrative examples.
\end{abstract}

\maketitle




\section{Introduction}\label{sec1}

The nonequilibrium nature of a dynamical system in interaction with external forces can be characterised by the statistics of the fluctuating (or stochastic) entropy production rate\newtext{~\cite{EvansPRL1993,EvansPRE1994,GallavottiPRL1995,GallavottiJSP1995,tasaki2000statistical,CrooksJSP1998,Crooks99,CollinNATURE2005}}. Its average is pivotal in understanding macroscopic properties of the system under study. If it is different from zero (i) the system is out of equilibrium, i.e., characterised by non-zero fluxes of energy between the system itself and the environment~\cite{EspositoNJP2010}, and (ii)
%
%
the arrow of time of the microscopic trajectories of the system cannot be reverted, viz.\ forward and backward processes are characterised by different dynamics~\cite{campisi2011fluctuation}.


\deletetext{The asymmetry of time can be understood in probability terms and is also derived in two established formalisms in nonequilibrium statistical physics, both involving the statistics of the fluctuating entropy production rate. }\deletetext{The first, in chronological order, is the one formulated by Gallavotti, Cohen, Evans, and Morriss (GCEM)~\cite{EvansPRL1993,EvansPRE1994,GallavottiPRL1995,GallavottiJSP1995}, which identifies an inherent symmetry in the entropy statistics that can be explained by properly weighting rare trajectories in the entropy production distribution. The other formalism is the one formulated by Tasaki and Crooks~\cite{tasaki2000statistical,CrooksJSP1998,Crooks99,CollinNATURE2005} that describes entropy production fluctuations through measurements of the work done on a system driven far from equilibrium. Tasaki and Crooks also identify an interesting symmetry, quantified by the ratio of the probability measures over the forward and backward trajectories in terms of the detailed balance equation.}
%
As shown in~\cite{LebowitzJSP1999,Barato2015}, at the microscopic level where fluctuations play a relevant role, the fluctuating entropy production of a dynamical process, namely $\mathbf{Z}$, modeling a physical system is provided by the action functional $\mathbb{S}$ via the Crooks' equality
\begin{equation}\label{eq:ActionFunctional_general_case}
    \exp\left(-\mathbb{S}\right) := \frac{\mathbb{P}_{\rm b}\left(R\left[ \mathbf{Z}\right]\right)}{\mathbb{P}_{\rm f}\left(\mathbf{Z}\right)}\,,
\end{equation}
where $R$ is a time-reversal operator, and $\mathbb{P}_{\rm f}$ and $\mathbb{P}_{\rm b}$ are the path-probability measures for the forward and backward processes. This interpretation of the action functional $\mathbb{S}$ as the fluctuating total entropy production stands only if $\mathbb{P}_{\rm f}$ and $\mathbb{P}_{\rm b}$ account for the initial conditions of the underlying dynamical process and, in particular, only if the initial condition of the backward process is obtained from the propagation~\footnote{This is given by applying the probability propagator of the process onto the initial distribution of the forward process.} of the initial condition of the forward process. Consequently, the fluctuating total entropy production will be composed by the sum of two terms: a first term accounting for the initial conditions, which is interpreted as the entropy production of the system itself, and a term involving the probability of forward and backward paths, which is interpreted as the entropy production originated by the interaction of the system with the environment.

%
%
\deletetext{It has been verified that} \newtext{As a consequence of \eqref{eq:ActionFunctional_general_case},} the expected value $\mathbb{E}_{\mathbb{P}_\text{f}}$ of $\exp\left(-\mathbb{S}\right)$ built over the forward path-probability measure $\mathbb{P}_\text{F}$ is always equal to $1$~\cite{Barato2015}, i.e.,
\begin{equation}\label{eq:Crooks_FT}
\mathbb{E}_{\mathbb{P}_\text{f}}[\exp\left(-\mathbb{S}\right)]=1 \,.
\end{equation}
This result is usually denoted as Integral Fluctuation Theorem for entropy production (IFT)\deletetext{. It is connected with the Jarzynski identities $\mathbb{E}_{\mathbb{P}_\text{F}}[\exp\left(-\beta\mathbb{W}\right)]=\exp\left(-\beta\Delta F\right)$~\cite{Jarzynski97} and $\mathbb{E}_{\mathbb{P}_\text{F}}[\exp\left(-\Delta\beta\,\mathbb{Q}\right)]=1$~\cite{Jarzynski04}, respectively for the statistics of work $\mathbb{W}$ done by the system and for the heat exchanged $\mathbb{Q}$ between two bodies kept at different temperatures ($\beta$ represents the inverse temperature).}\newtext{\footnote{It is connected with the Jarzynski identities $\mathbb{E}_{\mathbb{P}_\text{F}}[\exp\left(-\beta\mathbb{W}\right)]=\exp\left(-\beta\Delta F\right)$~\cite{Jarzynski97} and $\mathbb{E}_{\mathbb{P}_\text{F}}[\exp\left(-\Delta\beta\,\mathbb{Q}\right)]=1$~\cite{Jarzynski04}, respectively for the statistics of work $\mathbb{W}$ done by the system and for the heat exchanged $\mathbb{Q}$ between two bodies kept at different temperatures ($\beta$ represents the inverse temperature).}} \deletetext{Remarkably, these relations can be experimentally verified} \newtext{and it has been experimentally tested} for many physical systems working in fluctuating environments~\cite{Hummer2001,Liphardt2002,douarche2005experimental,HarrisPRL2007,Hummer2010,SairaPRL2012}. 


In this paper, we focus on discrete-time Markov processes and discuss a statistical estimator for the IFT in \eqref{eq:Crooks_FT} that can be evaluated in any numerical simulation and experimental protocol. This is obtained by replacing the expectation in \eqref{eq:Crooks_FT} with the sample mean (or empirical mean) evaluated over $N$ realisations (or copies) of the underlying dynamical process. While it is already well known that the IFT is not recovered due to convergence problems yielded by hard sampling of atypical (dominant) contributions to the sample mean~\cite{Gore2003,Jarzynski2006,Tietz2006,Speck2007,KimPRE2012comparison,SuarezPRE2012phase,HalpernPRE2016,HoangSciRep2016} and, possibly, the presence of \textit{absolute irreversibility}~\cite{MurashitaPREnonequilibrium}, here we provide an in-depth study of the IFT estimator by looking at microscopic Markov processes. 
%
%
In Section \ref{sec:breaking_FT}, via large deviation arguments we discuss the problem of directly, e.g., via Monte Carlo, `homogeneously' sampling the trajectory space of a dynamical system. For reversible processes, the probability to sample a forward and a backward trajectory with the same path probability, i.e., $\mathbb{S}=0$, asymptotically concentrates around $1$, thus fully recovering IFT for equilibrium systems. For nonequilibrium systems instead, $\mathbb{S}=0$ is naturally considered a rare event and its probability takes a large deviation form that scales exponentially with the time \deletetext{length} \newtext{duration} of the system. As a consequence, sampling rare (e.g., reversible) trajectories in nonequilibrium systems carries an exponential complexity in the number of samples that are needed to carefully cover the trajectory space guaranteeing convergence of the IFT estimator to $1$.
In Section \ref{sec:Estimate} we show a numerical recipe---formerly discussed in \cite{Rohwer2015} and applied here for the first time to the fluctuating entropy production rate of Markov processes---that allows one to properly set up an experimental protocol or a numerical simulation showing good convergence of the statistical estimator to IFT. Furthermore, for the particular case of Markov chains that quickly decorrelate over time, we show that the IFT estimator can be greatly improved enlarging its convergence region. \newtext{That decorrelation over time improves convergence of Markov-chain estimators is a result already known in the literature~\cite{Duffy2005,Rohwer2015,Sokal1997}, but here we exploit it in a physical context, improving convergence of the IFT estimator}. We corroborate our study with two illustrative examples of, respectively, an i.i.d.\ and a three-state Markov chain showing that estimates of the relevant parameters well match numerical simulations. We conclude in Section \ref{sec:concl} with a discussion of the results obtained, open questions, and ideas for future works.
%
%
%

%
%

\section{Model and sampling limits}
\label{sec:breaking_FT}

We consider a discrete-time~\footnote{This is to avoid a heavier (more technical) presentation of the results, as also done in the seminal paper of Crooks~\cite{CrooksJSP1998}} homogeneous Markov process $\mathbf{Z}_{n} = \left\lbrace Z_{0},Z_{1},\cdots,Z_{n} \right\rbrace$ with $Z_{n} \in \Gamma$ and $\Gamma$ a discrete state space. The dynamics of the Markov chain is governed by the transition matrix $\Pi$. In addition, we only consider ergodic Markov processes which admit a unique stationary distribution.  

We focus on the action functional introduced above and formalised as
\begin{equation}
    \label{eq:ActionFunctional}
    \begin{split}
    \mathbb{S}_{n} &\coloneqq - \ln \left( \frac{\mathbb{P}_{\Pi,\mu_0^b,n} \left( R \left[ \mathbf{Z}_{n} \right] \right)}{\mathbb{P}_{\Pi,\mu_0^f,n} \left( \mathbf{Z}_{n} \right)} \right)  \\
    &= - \ln(\mu_0^{b}(Z_{n})) + \ln(\mu_0^f(Z_{0})) + \sum_{\ell=0}^{n-1} \ln \left( \frac{\Pi \left( Z_{\ell},Z_{\ell+1} \right) }{\Pi \left( Z_{\ell+1}, Z_{\ell} \right) } \right) \ ,
    \end{split}
\end{equation} 
where $\mathbb{P}_{\Pi,\mu_0^f,n}$ is the forward path-probability measure of $\mathbf{Z}_{n}$ obtained by starting from the initial distribution $\mu_0^f$ and $\mathbb{P}_{\Pi,\mu_0^b,n}$ is the backward path-probability measure for the path of the reverse process (starting from $\mu_0^b$), with $R$ denoting the time-reversal operator as in Sect.~\ref{sec1}, see~\cite{LebowitzJSP1999,Barato2015} for further details. Furthermore, notice that the second equality of \eqref{eq:ActionFunctional} descends from the Markovianity property of both forward and backward path-probability measures, i.e.,
\begin{equation}
    \mathbb{P}_{\Pi,\mu_0^f,n}( \mathbf{Z}_n ) = \mu_0^f(Z_0)\prod_{\ell=0}^{n-1}\Pi(Z_\ell,Z_{\ell+1}) \ .
\end{equation}
As also pointed out in~\cite{Barato2015}, and previously mentioned in Sect.\ \ref{sec1}, depending on the choice of $\mu_0^b$, the action functional $\mathbb{S}_{n}$ can have different interpretations. In particular, in case the Markov process models a physical system and $\mu_0^b$ is simply defined as a propagation of the initial condition $\mu_0^f$ over the trajectorial probability measure of the process, $\mathbb{S}_n$ is interpreted as the fluctuating total entropy production of the Markov process. The reader should have in mind exactly this, in the following, every time we refer to $\mathbb{S}_n$ as an entropy production.

It is known~\cite{LebowitzJSP1999,Barato2015} that the
%
%
IFT in \eqref{eq:Crooks_FT} is satisfied for any $n$, provided that the expectation is taken over the forward path-probability measure.
%
%
Clearly, this is in agreement with Crooks'~\cite{CrooksJSP1998,Crooks99} and Jarzynski's~\cite{Jarzynski97} results, as argued in Sect.~\ref{sec1}. However, it is also worth observing that if we considered an action functional without initial conditions, then the corresponding fluctuation theorem of the form $\mathbb{E}_{\mathbb{P}_{\rm f}}[\exp\left(-\mathbb{S}_n\right)]=1$ would not be satisfied at every finite $n$, but only for $n \rightarrow \infty$~\cite{LebowitzJSP1999}.
 
Numerically (or experimentally), in order to carry out the expectation entering IFT, we need to run many independent realisations (labelled by the index $i=1,\ldots,N$) of the same stochastic process $\mathbf{Z}_n$. The IFT estimator thus reads
\begin{equation}
    \label{eq:CFTEstimator}
    \hat{G}_{N,n}^{\text{IFT}} \coloneqq \frac{1}{N} \sum_{i=1}^N e^{- \mathbb{S}_n^{(i)}} \ .
\end{equation}
Evidently, the IFT estimator depends on $n$ and $N$ that respectively characterise the extensivity of the action functional observable in \eqref{eq:ActionFunctional} and the number of independent realisations entering the estimate. It is well known, see~\cite{Gore2003,Duffy2005,Jarzynski2006,Speck2007,KimPRE2012comparison,SuarezPRE2012phase,Rohwer2015,HalpernPRE2016}, that direct sampling IFT by using \eqref{eq:CFTEstimator} may be a daunting computational problem which involves the correct sampling of rare events. In order to understand this argument we see in the following that it is useful to think of how equilibrium, or time-reversible, process and non-equilibrium processes behave in terms of entropy production. 

For both equilibrium and nonequilibrium processes IFT in \eqref{eq:Crooks_FT} holds, however the estimator in \eqref{eq:CFTEstimator} is strongly influenced by the process inherent nature. If $\mathbf{Z}_n$ is an equilibrium process, the probability for the action functional $\mathbb{S}_n$ to be zero, i.e., to sample a forward and a backward trajectory with the same probability, asymptotically concentrates around $1$ (exponentially quickly with $n$), thus fully recovering IFT. On the other hand, as soon as the system under investigation is out of equilibrium, the realisation $\mathbb{S}_n=0$ is a rare event as, naturally, the probability of a backward trajectory is always smaller than the probability of a forward one. Consequently, it is only by carefully sampling rare events that \eqref{eq:CFTEstimator} can converge to $1$, otherwise as $\mathbb{S}_n$ is typically greater than $0$ for a nonequilibrium process, we will have $\hat{G}_{N,n}^{\text{IFT}} < 1$.

Asymptotically with $n$, the probability distribution of $\mathbb{S}_n/n$ takes the large-deviation form
\begin{equation}
    \label{eq:LDPActFunc}
    \mathbb{P} \left( \frac{\mathbb{S}_n}{n} = s \right) \approx e^{-n I(s)} \ ,
\end{equation}
where the large-deviation rate function $I$ appears at exponent and characterises the likelihood of rare events. The rate function $I$ can be obtained via the following Legendre--Fenchel transform:
\begin{equation}
    \label{eq:LDFuncActFunc}
    I(s) = \sup_{k} \left( k s - \Psi(k) \right) \ ,
\end{equation}
where $\Psi(k)$ is the so-called Scaled Cumulant Generating Function (SCGF) defined as
\begin{equation}
    \label{eq:SCGF}
    \Psi(k) \coloneqq \lim_{n \rightarrow \infty} \frac{1}{n} \log G_{n}(k) \, ,
\end{equation}
where
\begin{equation}
    \label{eq:MomentActFunct}
    G_n(k) := \mathbb{E}_{\mathbb{P}_{\rm f}}\left[ e^{k \mathbb{S}_n} \right]  \ 
\end{equation}
is the action functional moment generating function. For details on large deviation theory we refer the reader to~\cite{DenHollander2000,Touchette2009,Dembo2010} and for methods to calculate \eqref{eq:MomentActFunct} to~\cite{LebowitzJSP1999,Barato2015,Touchette2018,Carugno2022}, just to mention a few works. 

Without having any a-priori knowledge on the underlying process $\mathbf{Z}_n$, one can readily observe that the exponential observable $\exp(-\mathbb{S}_n)$ concentrates (asymptotically with $n$) in probability around only three possible values: $0$, $1$, or $\infty$, respectively for $\lim_{n \rightarrow \infty} \mathbb{S}_n >$, $=$, or $<0$. First, we notice that if $\mathbf{Z}_n$ is an equilibrium process, $I(s)$ appearing in \eqref{eq:LDPActFunc} (and \eqref{eq:LDFuncActFunc}) has minimum (equal to zero) in $s=0$ which, for a physical system, corresponds to a null entropy production. The value $s=0$ characterises the law of large numbers, viz.\ in the thermodynamic limit $n \rightarrow \infty$ the observable $\mathbb{S}_n$ concentrates around $0$. Therefore, $\lim_{n \rightarrow \infty} e^{-\mathbb{S}_n} = 1$ in probability and, as a consequence, however big is the size of the sample statistics, IFT is always satisfied. In the opposite case, if $\mathbf{Z}_n$ is a nonequilibrium process, the large-deviation rate function is zero at $s>0$, physically corresponding to a non-zero entropy production. This is because a backward trajectory is typically less probable than the corresponding forward one as implied by the second law of thermodynamics. Consequently, $\lim_{n \rightarrow \infty} e^{-\mathbb{S}_n} = 0$ in probability and therefore the IFT estimator in \eqref{eq:CFTEstimator} converges to $0$ regardless of the sampling size. One may look at this as a statistical breaking of the estimator in the limit of infinitely long trajectories of nonequilibrum systems. The physical meaning of taking such a limit is open to debate. 

Numerically or experimentally, in order to be able to recover IFT, we need to make sure that we fully sample all rare events~\cite{Jarzynski2006,HalpernPRE2016}, and in particular those for $s \leq 0$ which have a backward trajectory more probable than a forward one. Unfortunately, these rare events are realised with an exponentially small probability, see \eqref{eq:LDPActFunc}, which makes it hard (impossible for $n \rightarrow \infty$) for the IFT estimator to sum up to $1$. One may be wrongly induced to conclude that to numerically obtain a value for the IFT estimator close to $1$, it should be enough to have an exponentially big sample size of the order $e^{n I(s=0)}$. Nevertheless, this turns out to not be sufficient as all contributions to the sum in \eqref{eq:CFTEstimator} have the same weight; hence, sampling just a few `good' rare events is not enough to tilt the empirical average towards $1$. Only a complete statistics of all rare events at finite $n$, which may very hard to achieve in practice for big $n$, should be able to recover IFT.

In the next Section, we further investigate numerical aspects related to the IFT statistical estimator and provide the reader with a recipe to recover the IFT in a numerical or experimental setting.

\section{Estimating IFT}
\label{sec:Estimate}

In this Section, we show how one could estimate IFT in a numerical or experimental setting when dealing with finite-time-$n$ discrete Markov processes and a finite sample size $N$. Specifically, we discuss a numerical recipe, formerly introduced in \cite{Rohwer2015}, for the observable $\mathbb{S}_n$ that allows one to properly set up an experiment without incurring in sampling problems. Furthermore, based on~\cite{Duffy2005}, we propose a technique that exploits the finite correlation time of the Markov chain to gain better convergence for the IFT statistical estimator.

In practice, one can estimate the moment generating function \eqref{eq:MomentActFunct} by sampling $N$ i.i.d.\,observations of $\mathbb{S}_n$ and evaluating
\begin{equation}
    \label{eq:EstMoment}
    \hat{G}_{N,n}(k) := \frac{1}{N} \sum_{i=1}^N e^{k \mathbb{S}_n^{(i)}} \ .
\end{equation}
Equivalently, one can estimate the SCGF $\Psi(k)$ in \eqref{eq:SCGF} by using
\begin{equation}
    \label{eq:SCGFEst}
    \hat{\Psi}_{N,n}(k) \coloneqq \frac{1}{n} \ln \hat{G}_{N,n}(k) \ .
\end{equation}
The estimate of the IFT as in \eqref{eq:CFTEstimator} is then obtained by replacing $k=-1$ in \eqref{eq:EstMoment}. 
Both estimators \eqref{eq:EstMoment} and \eqref{eq:SCGFEst} are known to be biased for finite $N$ due to downsampling of distribution tails of $\mathbb{S}_n$. This gives rise to \textit{linearisation} effects in the SCGF estimator in \eqref{eq:SCGFEst}, viz.\ when the sum of exponentials in \eqref{eq:EstMoment} is dominated by the largest sampled element for $k \rightarrow \infty$ (or smallest for $k \rightarrow -\infty$) tails of \eqref{eq:SCGFEst} become linear in $k$. For a comprehensive treatment of the problem, we refer the reader to \cite{Rohwer2015}. In the following, we borrow ideas from \cite{Rohwer2015} to establish when \eqref{eq:EstMoment} can be considered a good estimator of \eqref{eq:MomentActFunct}.

We start by remarking that, in order to have a good estimate of \eqref{eq:MomentActFunct}, we only need an accurate estimate of $\mathbb{P}(\mathbb{S}_n=ns) \equiv \mathbb{P}_n(\tilde{s})$, with $\tilde{s}=ns$, in a tiny region around the point $\tilde{s}^*(k)$. This point can be easily characterised by taking a Laplace approximation (for large $n$) of \eqref{eq:MomentActFunct}, i.e.,
\begin{equation}
    \label{eq:LaplaceApproxG}
    G_n(k) = \int_{-\infty}^{+\infty} ds \, e^{n k s} \mathbb{P}_n(\tilde{s}) \approx e^{ k \tilde{s}^*(k) + \ln \mathbb{P}_n(\tilde{s}^*(k)) } \ ,
\end{equation}
and hence by solving the following Euler--Lagrange equation~\footnote{Under the Laplace approximation, the Euler--Lagrange equation (\ref{eq:EulerLagrange}) is obtained by making the derivative of the exponent in the right-hand-side of Eq.~(\ref{eq:LaplaceApproxG}) with respect to $\tilde{s}^*$.}: 
\begin{equation}
    \label{eq:EulerLagrange}
    k \mathbb{P}_n(\tilde{s}^*) + \mathbb{P}'_n(\tilde{s}^*) = 0 \ ,
\end{equation}
where $(\cdot)'$ denotes the derivative with respect to $s$. Therefore, once again, if one has a good statistical sampling around $\tilde{s}^*(k)$, $G_n(k)$ is well estimated by $\hat{G}_{N,n}(k)$. In order to get such a good sampling, we require that with a very tiny probability that scales with $N$ a realization of $\mathbb{S}_n$ falls outside the region defined by $\left[ \bar{s}, +\infty \right]$ ($\left[ -\infty, \bar{s} \right]$) for the left-tail (right-tail) distribution, and with a constant probability $e^{-\tau}$ ($\tau$ is an arbitrarily small parameter) the observation falls inside that region. The end-point $\bar{s}$ of the convergence interval is implicitly defined as
\begin{equation}
    \label{eq:EndPointDef}
    F_n(\bar{s}) = \frac{\tau}{N} \ ,
\end{equation}
where $F_n$ is the cumulative distribution of $\mathbb{P}_n(\tilde{s})$, i.e.,
\begin{equation}
    \label{eq:CumulativeDistribution}
    F_n(s) := \int_{-\infty}^s d\tilde{s} \, \mathbb{P}_n(\tilde{s}) \ .
\end{equation}

In other words, for negative $k$ (but a specular argument holds for positive $k$ as well) if $\tilde{s}^* > \bar{s}$, then the sampling can be considered `good' and \eqref{eq:EstMoment} is a good estimate of \eqref{eq:MomentActFunct}. However, if $\tilde{s}^* < \bar{s}$, the sampling is `bad' and causes linearisation effects, and we need to increase its size in order to make $\tilde{s}^*$ fall inside the almost-sure sampled region. Hence, to find how the threshold $k_c$ moves as a function of the trajectory length $n$ and the sample size $N$ the following equality has to be solved:
\begin{equation}
    \label{eq:Threshold}
    \tilde{s}^*(k_c) = \bar{s}(n,N) \ .
\end{equation}
The solution $k_c(n,N)$ of the above equation allows one to set $N$ and $n$ in a numerical or experimental setting such that $k_c < - 1$, guaranteeing good convergence of the statistical estimator \eqref{eq:CFTEstimator} to IFT. 

It is worth noticing that since the observable action functional $\mathbb{S}_n$ has an additive structure, i.e., it is extensive in time, both $\tilde{s}^*$ and $\bar{s}$ are too. This extensivity property, as previously mentioned, may play a detrimental role in the convergence of the moment generating function estimator \eqref{eq:EstMoment} and consequently also for the IFT estimator \eqref{eq:CFTEstimator}. As a matter of fact, if the saddle point $\tilde{s}^*$ grows more rapidly with $n$ than $\bar{s}$ does, it will certainly be very hard to set up a good sampling window around $\tilde{s}^*$ for long trajectories. 

Let us observe that for general dynamical processes in which correlation times are hard to find or estimate, one is solely bound to use \eqref{eq:Threshold} to understand whether and how the IFT estimator shows good convergence. However, for Markov processes that are mixing in time, i.e., they `quickly' decorrelate~\footnote{The mathematical definition of mixing is rather technical. We refer to~\cite{Arnold1989} for further details.} with a finite correlation \deletetext{length} \newtext{time}~\footnote{We also remark that having a finite correlation \deletetext{length} \newtext{time} does not imply that the underlying process is time reversible; it is indeed possible to have non-equilibrium systems that have finite correlation \deletetext{lengths} \newtext{times}. We show an example in the following.}, we show in the following that the dependence on $n$ of $\mathbb{S}_{n}$ can be ruled out allowing for a much better convergence of the IFT estimator. In this regard, let us suppose we know or can estimate the finite correlation \deletetext{length} \newtext{time} \deletetext{$\tau$}\newtext{$\xi$} of the Markov process $\mathbf{Z}_n$. Assuming $n$ multiple of \deletetext{$\tau$}\newtext{$\xi$} and $\eta = n/\deletetext{\tau}\newtext{\xi}$, Eq.~\eqref{eq:ActionFunctional} can be rewritten as
\begin{equation}
    \label{eq:ActionFunctionalCorr}
    \mathbb{S}_n = - \ln \left( \mu_0^b(Z_n) \right) + \ln \left( \mu_0^f(Z_0) \right) + \sum_{\ell=0}^{\eta} Y_{\ell} \ ,
\end{equation}
where
\begin{equation}
    \label{eq:YellClustered}
    Y_{\ell} = \sum_{\ell' = 0}^{\deletetext{\tau}\newtext{\xi}-1} \ln \left( \frac{\Pi(Z_{\deletetext{\tau}\newtext{\xi} \ell+\ell'},Z_{\deletetext{\tau}\newtext{\xi} \ell+\ell'+1})}{\Pi(Z_{\deletetext{\tau}\newtext{\xi} \ell+\ell'+1},Z_{\deletetext{\tau}\newtext{\xi} \ell+\ell'})} \right) \ .
\end{equation}
Under ergodicity and the mixing condition for the Markov chain, in the limits $n \rightarrow \infty$ and $\deletetext{\tau}\newtext{\xi} \rightarrow \infty$ but with $n \gg \deletetext{\tau}\newtext{\xi}$, it can be shown that the variables $Y_\ell$ become independent and identically distributed~\cite{Duffy2005}. Therefore, the moment generating function \eqref{eq:MomentActFunct} can be simplified as follows:
\begin{equation}
    \label{eq:MomentActFunctCorr}
    G_n(k) \approx \left( \frac{\mu_0^f(Z_0)}{\mu_0^b(Z_n)} \right)^k \left( \mathbb{E}_{Y_\ell} \left[ e^{k Y_\ell} \right] \right)^\eta \ ,
\end{equation}
where the factorisation over $Y_\ell$ variables is consequence of Cram\'{e}r theorem~\cite{Cramer1938,Cramer2018}. The estimation problem is then reduced to properly sampling the probability distribution of $Y_\ell$. Hence, the estimator of the expectation appearing in \eqref{eq:MomentActFunctCorr} reads
\begin{equation}
    \label{eq:EstMomentCorr}
    \hat{G}_N(k) \coloneqq \frac{1}{N} \sum_{i=1}^N e^{k Y_\ell^{(i)}} 
\end{equation}
and the SCGF estimate in \eqref{eq:SCGFEst} simplifies to 
\begin{equation}
    \label{eq:SCGFEstCorr}
    \hat{\Psi}_N = \frac{1}{\deletetext{\tau}\newtext{\xi}}  \ln \hat{G}_N(k) \ , 
\end{equation}
whereby, evidently, there is no longer dependence on $n$, the \deletetext{length} \newtext{duration} of the process. The use of this technique is known to yield an exponential gain in estimation compared to estimating directly $\mathbb{S}_n$ as in \eqref{eq:EstMoment} or \eqref{eq:SCGFEst}, see~\cite{Duffy2005,Rohwer2015} for details. As the focus is now shifted onto the observable $Y_\ell$, by retracing the steps above we can get an equation that is equivalent to \eqref{eq:Threshold}. This reads
\begin{equation}
    \label{eq:ThresholdCorr}
    y^*(k_{c,y}) = \bar{y}(N) \ ,
\end{equation}
where $y^*(k_{c,y})$ and $\bar{y}$ are respectively the saddle point associated with the moment generating function and the end point of the cumulative distribution function of $Y_\ell$. Likewise before, \eqref{eq:ThresholdCorr} can be used to determine the critical value $k_{c,y}$ that bounds the convergence of the estimator (\ref{eq:EstMomentCorr}). As the length of the trajectory plays against the sampling accuracy, we must have 
\begin{equation}
    \label{eq:kcCorrBetter}
    |k_{c,y}| \geq |k_c| \ \deletetext{.} \newtext{,}
\end{equation}
\newtext{at fixed $N$ and for any $n \geq \deletetext{\tau}\newtext{\xi}$.} 

\newtext{We conclude this Section with an important remark on the methods investigated and for the following illustrative examples. We will always assume to know the stationary distribution of the Markov chain investigated and that will be used to initialise the Markov chain on the state space. By considering $\mu_0^b$ the natural time-propagation of the initial condition, and by definition of stationarity, $\mu_0^b$ will also be the stationary distribution. Therefore, boundary terms appearing in \eqref{eq:ActionFunctional} and \eqref{eq:ActionFunctionalCorr} cancel out. Furthermore, as we initialise the Markov chain in its stationary distribution, autocorrelation effects, which are related to the relaxation time of the Markov chain to its stationary state, leading to systematic sampling errors~\cite{Sokal1997} are not present. Clearly, if the stationary distribution was not known, one would need to introduce a further, independent, sampling procedure to estimate it. At the moment, this is out of scope of our work and we will not discuss it any further.}
\deletetext{In the following, we will show two examples where this approach makes the estimate much more precise.}


\section{Examples}
\label{ref:secex}

In the following, we investigate two simple examples of nonequilibrium systems: (i) an i.i.d.\,stochastic process, and (ii) a three-state Markov chain. In both cases, we show that numerically IFT is not always well estimated, and that by solving \eqref{eq:Threshold} we can determine how we need to scale the number of copies $N$ with the trajectory length $n$, or vice versa, in order to obtain a good convergence of the IFT estimator. Furthermore, we show that by knowing the correlation \deletetext{length} \newtext{time} of the system it is possible to design a better estimator for the full SCGF and, as a consequence, for the IFT too.

\subsection{Trick-coin tossing}

We consider a trick coin. A side of the coin is marked with $a$ and the opposite side with $c$. The probability of getting $c$ out of a single toss is $p \in [0,1]$, whereas the probability of getting $a$ is $1-p$. We focus on the observable $\mathbb{S}_n$ of Eq.~\eqref{eq:ActionFunctional} which, in this simplified scenario, can be rewritten as
\begin{equation}
    \label{eq:ThreeStateActionFunctional}
    \mathbb{S}_n = \left( 2 M_c - n \right) \ln \left( \frac{p}{1-p} \right) \ ,
\end{equation}
where $M_c$ is a random variable representing the number of $c$-outcome tosses.

It is immediate to verify that \eqref{eq:ThreeStateActionFunctional} follows a binomial probability distribution of the form
\begin{equation}
    \label{eq:ThreeStateBinDistr}
    \mathbb{P}_n(\mathbb{S}_n = \tilde{s}) \coloneqq \mathbb{P}_n(M_c=m_c) = {n \choose m_c} p^{m_c} (1-p)^{n-m_c} \ .
\end{equation}
Thus, the moment generating function \eqref{eq:MomentActFunct} and the SCGF \eqref{eq:SCGF} can be derived analytically, and they read as
\begin{align}
    \label{eq:ThreeStateMoment}
    G_n(k) &= p^{-k n} (1-p)^{n(1+k)} \left( 1 + \left( \frac{1-p}{p} \right)^{-2k-1} \right)^n \\
    \label{eq:ThreeStateSCGF}
    \Psi(k) &= \ln \left( p^{-k} (1-p)^{1+k} \left( 1 + \left( \frac{1-p}{p} \right)^{-2k-1} \right) \right) \ .
\end{align}
Moreover, we also note that by replacing $k=-1$ in \eqref{eq:ThreeStateMoment} (and \eqref{eq:ThreeStateSCGF}) we get $G_n(-1) = 1$ ($\Psi(-1)=0$). Hence, IFT is satisfied. 

In what follows, we show how one should set up relevant parameters, such as the number of simulations or experiments and their \deletetext{length} \newtext{duration}, to make the estimator \eqref{eq:EstMoment} properly converge to $1$ for $k=-1$. In order to do so, we start off by calculating $m^*(k)$, solution of the Euler--Lagrange equation \eqref{eq:EulerLagrange} for $\mathbb{P}_n(m_c)$. This, in the saddle-point approximation $n \gg 1$, is
\begin{equation}
    \label{eq:ThreeStateSaddle}
    m_c^*(k) = \frac{n}{\left( 1+ \left( \frac{1-p}{p} \right)^{1+2k} \right)} \ .
\end{equation}
As expected, if we replace $k=0$ in \eqref{eq:ThreeStateSaddle}, we get $m_c^*(k)=n p$, viz.\ the typical value of $\mathbb{P}_n( m_c)$ is $n p$. Now, we need to calculate $\bar{m}_c(n,N)$, end-point solution of \eqref{eq:EndPointDef}. We postpone the full calculation to the Appendix \ref{sec:appmc} and we show here only the result
\begin{equation}
    \label{eq:ThreeStateEndPoint}
    \bar{m}_c(n,N) = n p + \sqrt{ 2 n p (1-p) \ln N} \ ,
\end{equation}
which is valid for $p \sim O(1)$, $p<1/2$ and, once again, for $n \gg 1$. Notice that a similar derivation can be carried out for the case $p>1/2$. However, we do not need to do it as we know that, due to the inherent symmetry of the observable $\mathbb{S}_n$, $\Psi(k)$ is invariant under the transformation $p \rightarrow 1-p$.

Then, for this example, the equation \eqref{eq:Threshold} is 
\begin{equation}
    \label{eq:ThreeStateKcToStudy}
    p + \frac{\sqrt{2 p (1-p)}}{\sqrt{n}} \sqrt{\ln N} = \left( 1+ \left( \frac{1-p}{p} \right)^{1+2k_c} \right)^{-1} \ .
\end{equation}
This equation allows us to draw interesting conclusions on how we should set $n$ and $N$ in a numerical or lab experiment to have good convergence of the estimator $\hat{G}_{N,n}(k)$.

Asymptotically, from \eqref{eq:ThreeStateKcToStudy}, it is easy to show how the critical value of $k_c$ scales with $N$. In this regard, we remind the reader that if $k < |k_c|$ then $\hat{G}_{N,n}(k)$ converges in probability to $G_{n}(k)$, otherwise the estimator is biased~\cite{Rohwer2015}. Moreover, from \eqref{eq:ThreeStateKcToStudy} we immediately see that for $n \rightarrow \infty$, we have $k_c \rightarrow 0$, viz.\ in the thermodynamic limit $n \rightarrow \infty$ the estimator is not biased only for $k=0$. Furthermore, in the regime $\ln N \gg n$ and for $k < 0$, we have that $k_c \sim \ln ( \ln ( \ln ( N ) ) )$. Instead, if $\ln N \sim n$ or $\ln N \ll n$, $k_c \sim \text{constant}$. It is clear that even in the best sampling scenario, i.e., $\ln N \gg n$, $k_c$ scales very slowly with $N$. In practical applications where it is likely that the sampling size is smaller than an exponential in $n$, $k_c$ behaves as a constant. By rearranging \eqref{eq:ThreeStateKcToStudy} we can derive a general equation for the scaling of $k_c$ with $n$ and $N$. We get
\begin{equation}
    \label{eq:ThreeStateKcn}
    k_c(n,N) = -\frac{1}{2 \ln \left( \frac{1-p}{p} \right)} \left( -\ln \left( \frac{\sqrt{n}}{\sqrt{n}p + \sqrt{2p(1-p)\ln N}} - 1 \right) + \ln \left( \frac{1-p}{p} \right) \right) \ .
\end{equation}

\begin{figure}
    \centering
    \includegraphics[width=0.5\linewidth]{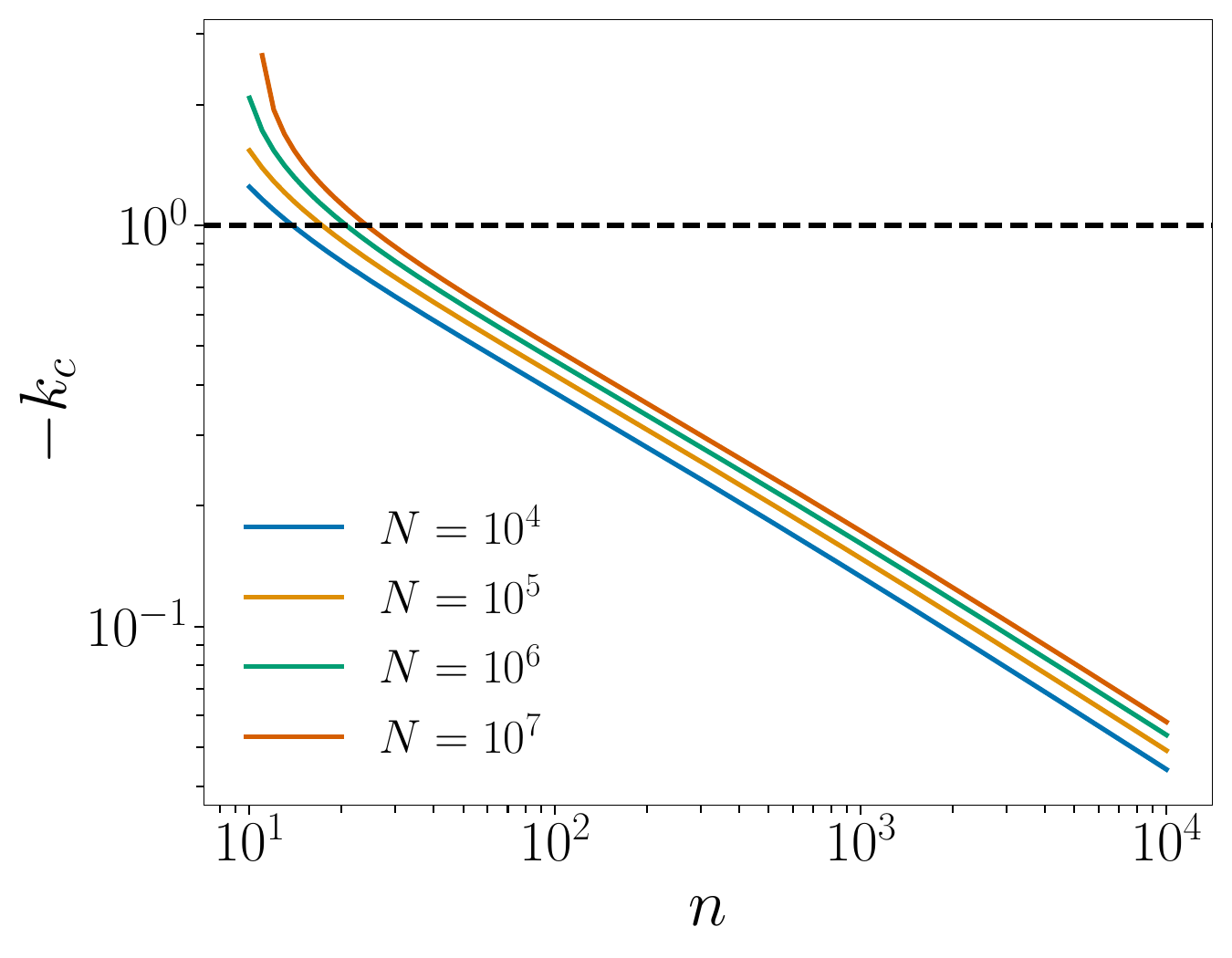}
    \caption{
    $k_c$ (coloured solid lines) as a function of $n$ in \eqref{eq:ThreeStateKcn} for different sampling sizes $N$ at fixed $p=0.25$ compared with $k_c=-1$ (black dashed line).
    }
    \label{fig:Scaling}
\end{figure}

In Fig.\ \ref{fig:Scaling} we show how $k_c$ scales with $n$ with varying sampling size $N$. We notice that, for $n \gg 1$, $k_c$ quickly converges to $0$. Furthermore, although the scaling is the same, the greater the $N$, the greater the $k_c$ in absolute value for a fixed value of $n$. In particular, since we are chiefly interested in the convergence of the IFT estimator we show how $k_c$ crosses $-1$. Obviously, as already mentioned, the longer the sampled trajectory, the bigger the sample size required to have good convergence to IFT. However, the scaling plays against numerics: it is clear that one needs to exponentially increase the sample size (from $N=10^4$ to $10^7$) to gain convergence to the IFT for slightly longer trajectories (from approximately $n=13$ to $30$ steps).

\begin{figure}
    \centering
    \includegraphics[width=0.5\linewidth]{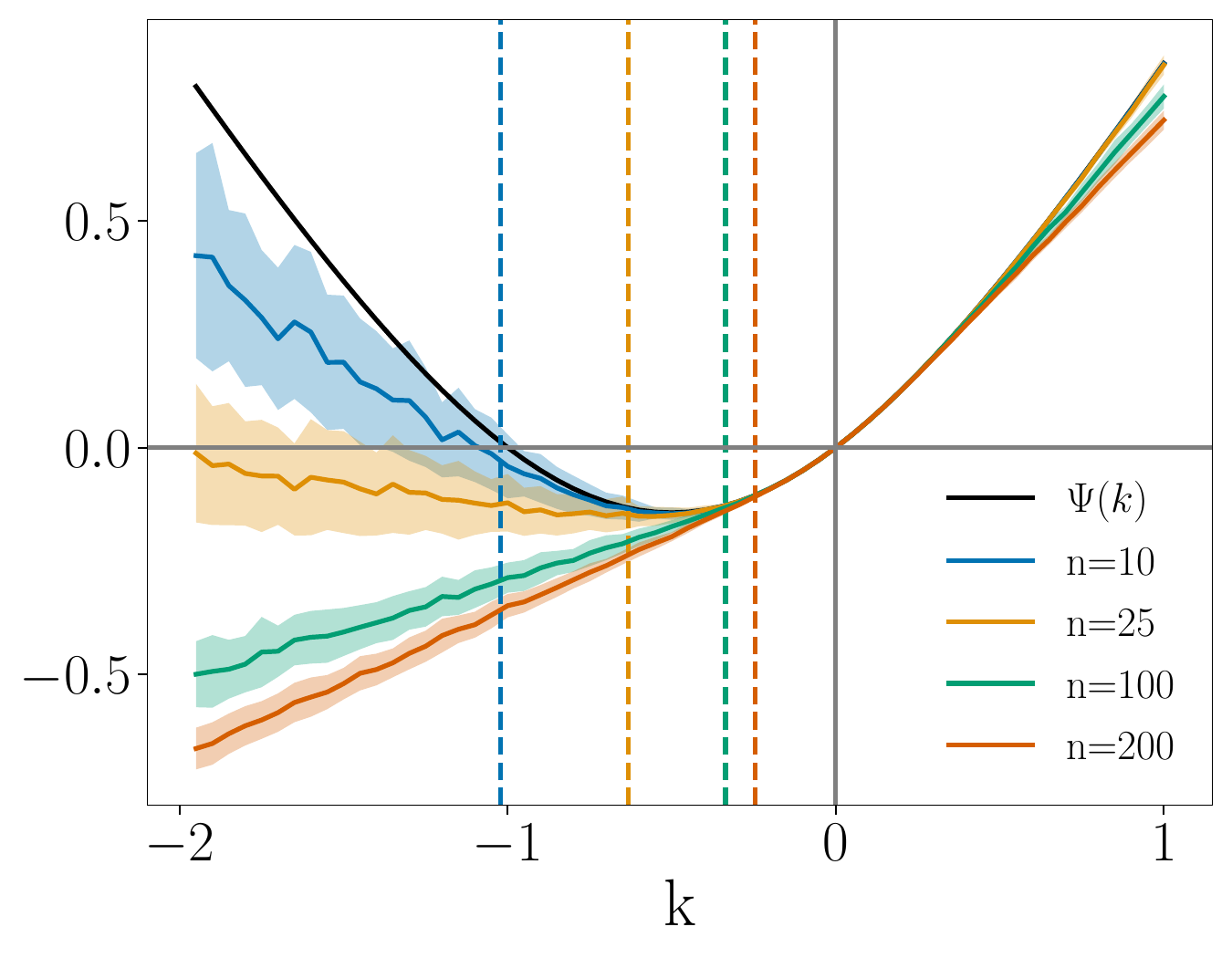}
    \caption{Estimated SCGFs $\hat{\Psi}_{N,n}(k)$ (coloured solid curves) obtained as in \eqref{eq:SCGFEst} for different values of $n$ with a sample size of $N=10^5$ copies (the shaded area represents a standard deviation error over $100$ samples). These curves are compared with the theoretical SCGF $\Psi(k)$ (solid black line) in \eqref{eq:ThreeStateSCGF} and with values of $k_c$ (coloured vertical dashed lines) with varying $n$.}
    \label{fig:scgf}
\end{figure}

In Fig.\ \ref{fig:scgf} we plot the estimated SCGFs given by \eqref{eq:SCGFEst} for different values of $n$ at fixed $N=10^5$, and we compare them with both the true SCGF $\Psi(k)$ (given by \eqref{eq:ThreeStateSCGF}) and vertical lines corresponding to critical values of $k_c$. We notice that estimates are good up to a certain value of $k$ after which linearisation effects, due to tail down-sampling, take place. Remarkably, for values of $n \gtrsim 25$, we can predict with great accuracy the values of $k_c$. This is clear by the fact that the vertical dashed lines cross the coloured solid lines very close to the point where these last depart from the true SCGF. Theoretically, IFT is satisfied at $k=-1$, i.e., $\Psi(-1)=0$. Estimates reveal that the larger the $n$, the worse is the convergence to IFT. In particular, with the sample size adopted here, it seems that only for $n=10$ we can have good convergence of the estimator to the IFT's result as $k_c \lesssim -1$. We also remark that beyond $k_c$ the moment generating function estimator is strongly biased, departing to negative values for $k<-1$ already for $n \gtrsim 25$.

\begin{figure}
    \centering
    \includegraphics[width=0.5\linewidth]{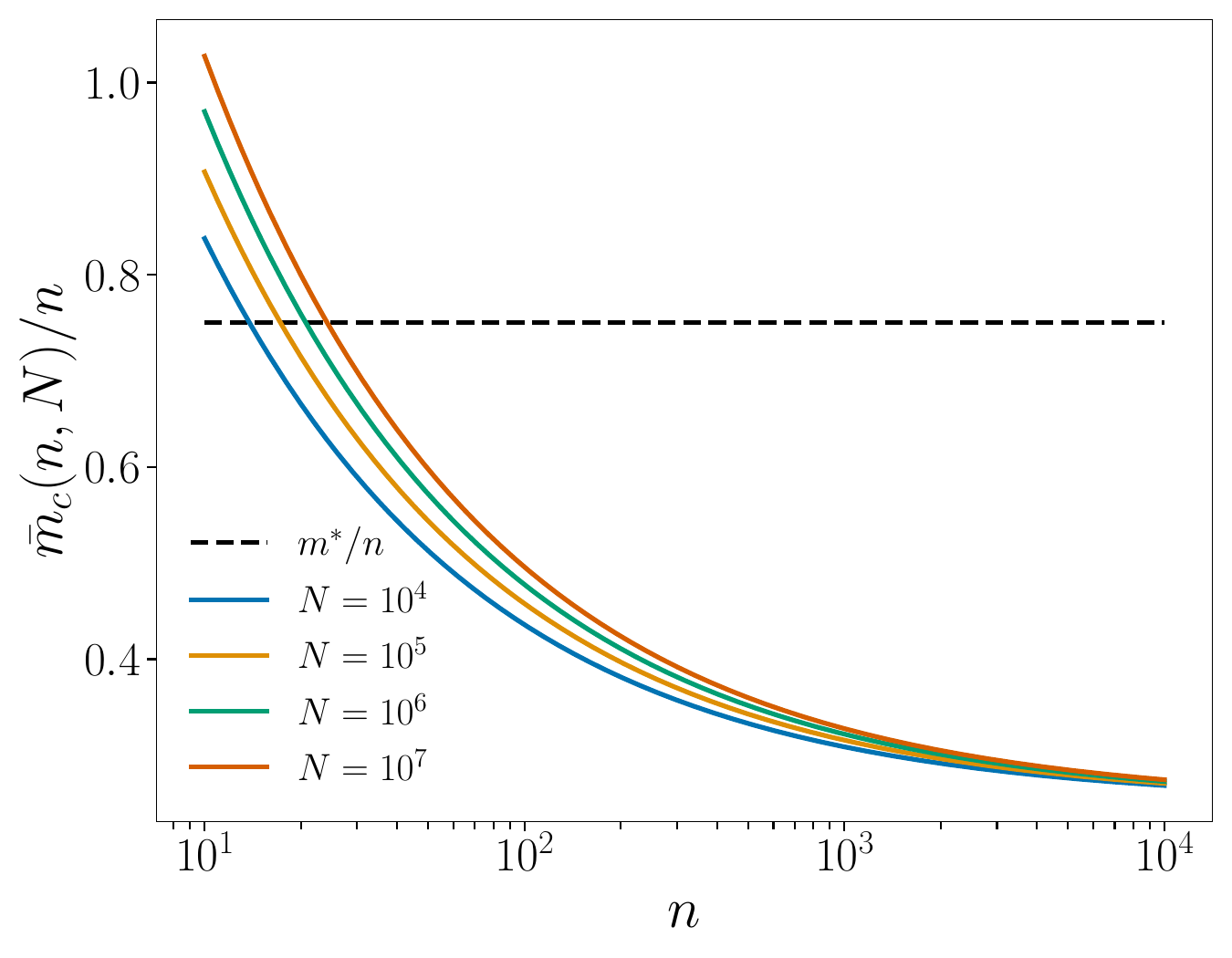}
    \caption{$m_c^*/n$ (black dashed line) in \eqref{eq:ThreeStateSaddle} and $\bar{m}_c/n$ (coloured solid lines) in \eqref{eq:ThreeStateEndPoint} at fixed $k=-1$, as a function of $n$ for various sample sizes $N$ and at fixed $p=0.25$.}
    \label{fig:ScalingFixedk}
\end{figure}

As previously mentioned, the bias in the curves of Fig.\ \ref{fig:scgf} is due to the different scaling of $k_c$ with respect to $n$. In Fig.\ \ref{fig:ScalingFixedk} we plot $m_c^*/n$ in \eqref{eq:ThreeStateSaddle} and $\bar{m}_c/n$ in \eqref{eq:ThreeStateEndPoint}, at fixed $k=-1$, as a function of $n$ for various sample sizes $N$. Evidently, the saddle point $m^*_c/n$ lies below $\bar{m}_c/n$ already for short trajectories, i.e., $n \lesssim 25$. Furthermore, as already noticed in Fig.\,\ref{fig:Scaling}, 
%
%
it is unlikely to enlarge the sampling window to include $m_c^*$ by varying the sample size $N$. 

Interestingly, much work has been done on understanding this bias for free-energy equilibrium estimates obtained from sampling non-equilibrium work values in physical settings~\cite{Wood1991,Hummer2001,Liphardt2002,Zuckerman2002,Zuckerman2002a,HarrisPRL2007,Hummer2010,Palassini2011}. However, it seems clear that to compensate the bias one needs to have a precise understanding of the distribution tails~\cite{Palassini2011}, but
%
%
this information may often not be available. Although this is not the case here (we have the full distribution of $\mathbb{S}_n$), for didactic purposes we show how we can get rid of the bias by using the method proposed at the end 
%
%
of Section \ref{sec:Estimate}, i.e., by estimating $G_n(k)$ with the new estimator \eqref{eq:EstMomentCorr} that exploits the fact that $\mathbf{Z}_n$ has a finite correlation \deletetext{length} \newtext{time}.

\begin{figure}
    \centering
    \includegraphics[width=0.5\linewidth]{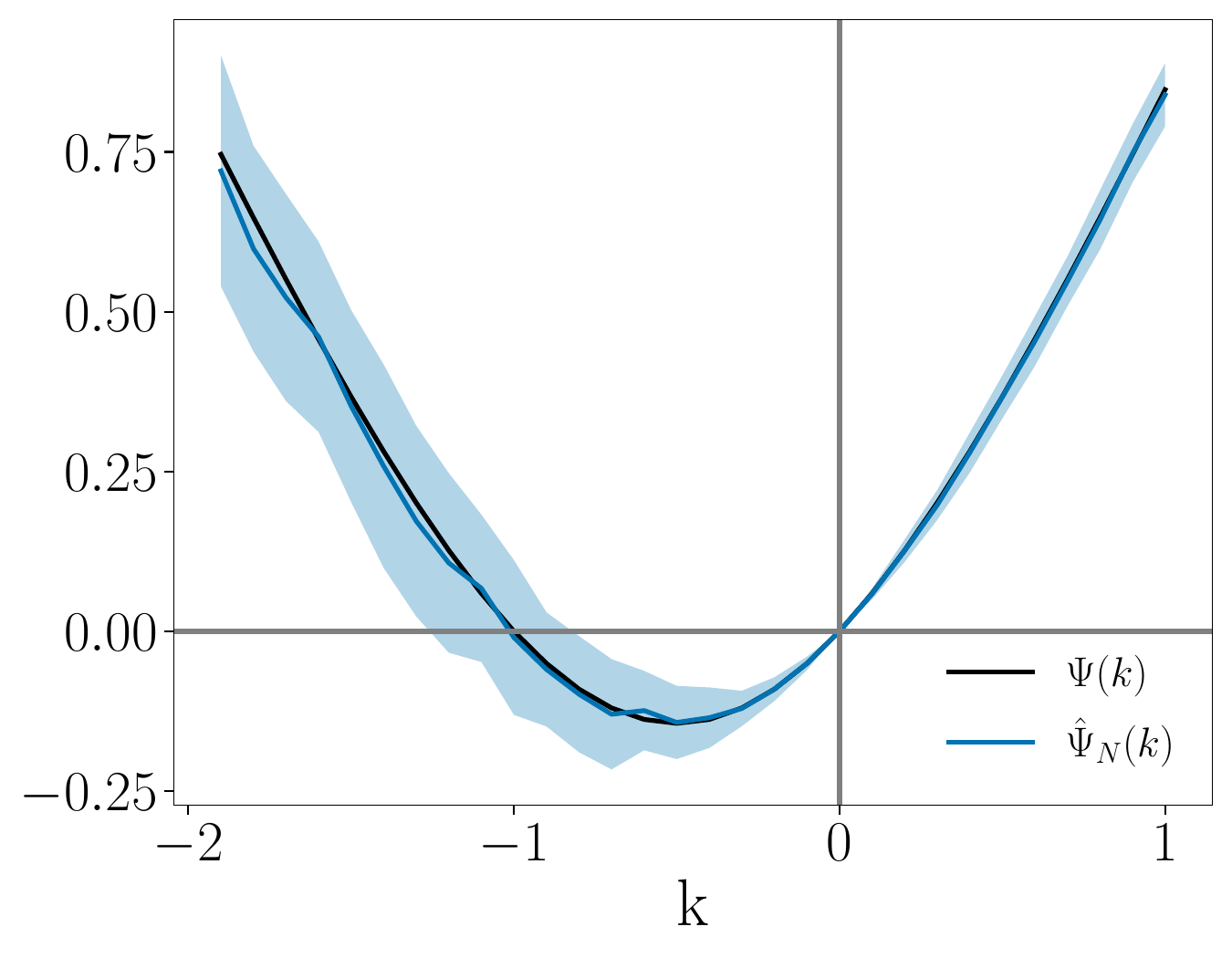}
    \caption{Estimated SCGF (blue solid curve) given by $\hat{\Psi}_N(k) = k \ln (1-p)/p + \ln \hat{G}_N(k)$ with $N=10^4$ copies (the shaded region corresponds to one standard deviation computed over $100$ samples) compared with the true SCGF $\Psi(k)$ (black solid curve).}
    \label{fig:scgfcorr}
\end{figure}

For the example of 
the trick-coin tossing investigated here, the correlation \deletetext{length} \newtext{time} is just \deletetext{$\tau=1$} \newtext{$\xi=1$} as we are dealing with an i.i.d.\ process. Thus, by replacing in \eqref{eq:MomentActFunct} the explicit form of $\mathbb{S}_n$ given by \eqref{eq:ThreeStateActionFunctional} and after carrying out some simple algebra, we obtain
\begin{equation}
    \label{eq:ThreeStateMomentActFuncCorr}
    G_{n}(k) = \left[ \frac{1-p}{p} \right]^{n k} \left( \mathbb{E}_{Y_\ell} \left[ \exp \left( 2 k Y_\ell \ln \frac{p}{1-p} \right) \right] \right)^n \ ,
\end{equation}
where we also made the substitution $M_c = \sum_{\ell=1}^n Y_\ell$ with $Y_\ell$ a Bernoulli random variable taking value $Y_\ell=1$, i.e., coin toss results in $c$, (or $Y_\ell=0$ meaning the coin toss results in $a$) with probability $p$ (or $1-p$). Remarkably, the estimation problem is now greatly simplified as we only need to sample a Bernoulli random variable. Then, the estimator \eqref{eq:EstMomentCorr} for this particular example reads
\begin{equation}
    \label{eq:ThreeStateMomentActFuncCorrEst}
    \hat{G}_N(k) = \frac{1}{N} \sum_{i=1}^N \exp \left( 2 k Y_i \ln \frac{p}{1-p} \right) \ .
\end{equation}
Noticeably, we passed from a random variable $\mathbb{S}_n$ that, although still bounded for finite $n$, becomes unbounded for $n \rightarrow \infty$, to a random variable $Y_\ell$ that is always bounded. In particular, $Y_\ell$ can take only two values and for this reason its probability distribution tails are well estimated as soon as these two values are sampled just once. As a consequence, $k_c \rightarrow \pm \infty$ and hence no linearisation effects appear. We plot the estimated SCGF given by $\hat{\Psi}_N (k) = k \ln (1-p)/p + \ln \hat{G}_N(k)$ and compare it with the true SCGF $\Psi(k)$ in Fig.\,\ref{fig:scgfcorr}. Remarkably, no bias nor linearisation effects take place already for $N=10^4$. Hence, we see a great improvement with respect to the estimates obtained out of \eqref{eq:EstMoment} in Fig.\,\ref{fig:scgf}.

\subsection{Three-state Markov chain}

We consider a discrete-time Markov chain evolving on a state space composed by three states $\left\lbrace -1, 0, 1 \right\rbrace$ all connected to each other. The transition matrix of the Markov chain is
\begin{equation}
    \label{eq:RealThreeStateTransMatrix}
    \Pi =
    \begin{pmatrix}
        0 & p_{-1} & 1-p_{-1} \\
        1-p_0 & 0 & p_0 \\
        p_1 & 1-p_1 & 0 \\
    \end{pmatrix} \ ,
\end{equation}
with independent probabilities $p_{-1},p_0,p_1 \in [0,1]$. [Notice that if we fix $p_{-1}=p_0=p_1=p$ we end up with the trick-coin example studied above. \newtext{Furthermore, although in this example we consider diagonal elements of $\Pi$ to be $0$, this is not a working assumption for the application of the discussed methods.}] The stationary probability distribution of the Markov chain, which will be useful later on, can easily be computed as the left eigenvector of $\Pi$ in \eqref{eq:RealThreeStateTransMatrix} and reads
\begin{equation}
    \label{eq:RealThreeStateStatDistr}
    \mu_0^f = 
    \begin{pmatrix}
        \frac{1+p_0(p_1-1)}{3+p_0(p_1-1)-p_1+p_{-1}(p_0+p_1-1)} \\
        \frac{1+p_1(p_{-1}-1)}{3+p_0(p_1-1)-p_1+p_{-1}(p_0+p_1-1)} \\
        \frac{1+p_{-1}(p_0-1)}{3+p_0(p_1-1)-p_1+p_{-1}(p_0+p_1-1)} \\
    \end{pmatrix} \ .
\end{equation}
We focus again on the observable $\mathbb{S}_n$ of Eq.~\eqref{eq:ActionFunctional} that can be rewritten as
\begin{equation}
    \label{eq:RealThreeStateActionFunctional}
    \mathbb{S}_n = n \left[ \ln \left( \frac{\Pi(-1,0)}{\Pi(0,-1)} \right) \mathbb{J}_{n}(-1,0) + \ln \left( \frac{\Pi(0,1)}{\Pi(1,0)} \right) \mathbb{J}_{n}(0,1) + \ln \left( \frac{\Pi(1,-1)}{\Pi(-1,1)} \right) \mathbb{J}_{n}(1,-1) \right] \ ,
\end{equation}
where we have also introduced the so-called current $\mathbb{J}_n(z,z')$ which is the net fraction of jumps of the Markov chain between state $z$ and $z'$.

\begin{figure}
    \centering
    \includegraphics[width=0.5\linewidth]{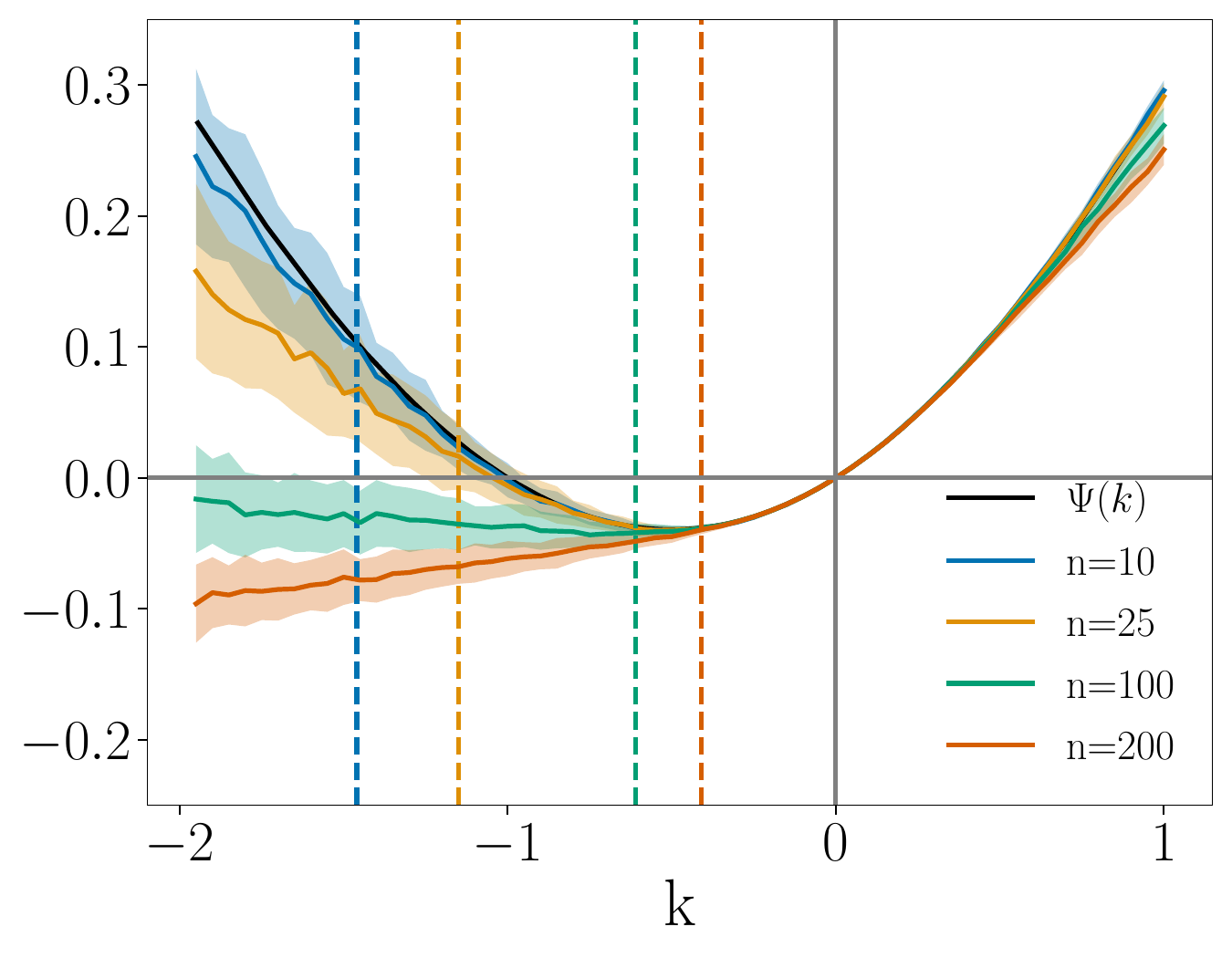}
    \caption{Estimated SCGFs $\hat{\Psi}_{N,n}(k)$ (coloured solid curves) obtained as in \eqref{eq:SCGFEst} for different values of $n$ with a sample size of $N=10^5$ copies (the shaded area represents a standard deviation error over $100$ samples). These curves are compared with the theoretical SCGF $\Psi(k)$ (solid black line) and with values of $k_c$ (coloured vertical dashed lines) with varying $n$. Simulations are run with $p_{-1}=0.5$, $p_0=0.4$, and $p_1=0.2$.}
    \label{fig:scgf3}
\end{figure}

As mentioned in Sect.\ \ref{sec:breaking_FT}, the SCGF $\Psi(k)$ can be calculated in different ways. Here, we make use of a method discussed in~\cite{Carugno2022}---see also \cite{Carugno2022a} for an application---but other methods could also be used, e.g., studying the dominant eigenvalue of the so-called tilted matrix~\cite{Touchette2018} associated with $\mathbb{S}_n$. The SCGF is plotted as a solid black line in Fig.\ \ref{fig:scgf3}, where it is easy to check that IFT is satisfied, i.e., $\Psi(-1)=0$.

\begin{figure}
    \centering
    \includegraphics[width=0.5\linewidth]{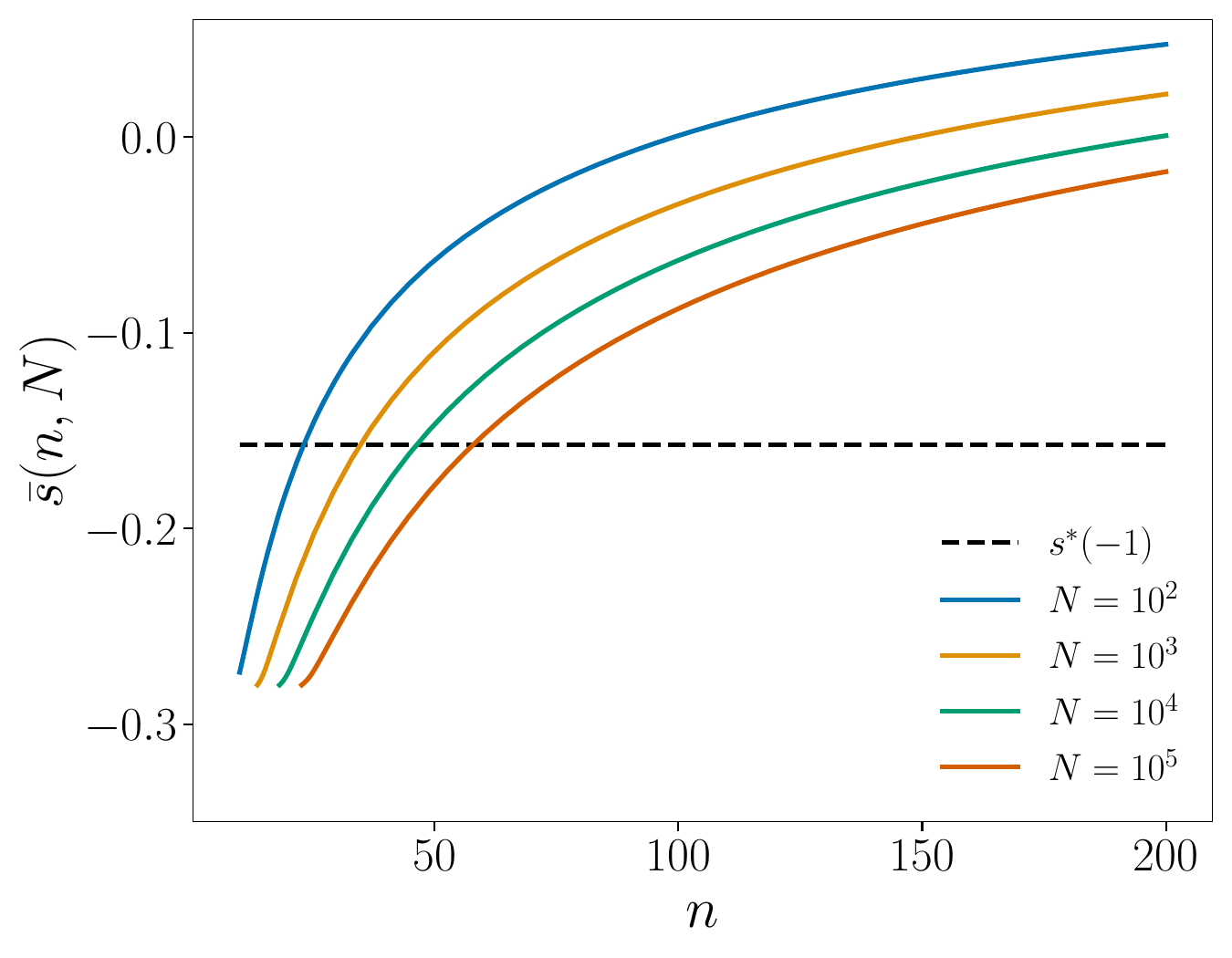}
    \caption{$\bar{s}(n,N)$ (coloured solid lines) with varying $N$ compared with the value $s^*(-1)$ (black dashed line).}
    \label{fig:sbar_scaling}
\end{figure}

The SCGF can directly be estimated by making use of \eqref{eq:EstMoment} which, for a fixed length $n$ of the process, samples $N$ values for the observable $\mathbb{S}_n$ in \eqref{eq:RealThreeStateActionFunctional}. As previously mentioned, to have a good statistical sampling of IFT we need $s^*(-1) \geq \bar{s}(n,N)$ (notice that differently from before, now $s^*$ and $\bar{s}(n,N)$ are already rescaled by $n$). On the one hand, the value $s^*$ can easily be calculated by making use of the Legendre duality relation
\begin{equation}
    \label{eq:RealThreeStateLegendreDuality}
    s^*(k) = \Psi'(k) \ .
\end{equation}
On the other hand, the value $\bar{s}(n,N)$ can be estimated for $n \gg 1$ from \eqref{eq:CumulativeDistribution} by replacing the large-deviation relation \eqref{eq:LDPActFunc} and by (numerically) inverting the rate function $I$. Eventually, we get
\begin{equation}
    \label{eq:RealThreeStateBars}
    \bar{s}(n,N) \approx I^{-1} \left( -\frac{1}{n} \ln \left( \frac{1}{N} \right) \right) \ ,
\end{equation}
where we imposed $\tau=1$ in \eqref{eq:EndPointDef}. As we are chiefly interested in the convergence of the IFT statistical estimator in \eqref{eq:CFTEstimator} we focus on the value $k=-1$ and plot in Fig.\ \ref{fig:sbar_scaling} $s^*(-1)$ as a black dashed line and $\bar{s}(n,N)$ as coloured curves for a few fixed values of $N$. Focusing on the line $s^*(-1)$, all lengths $n$ of the process that are found on the left of the coloured lines will give rise to a good convergence of the statistical estimator, for a sample size $N$ corresponding to that particular coloured line we are looking at. Evidently, the greater the $N$, the longer the \deletetext{length} \newtext{duration} of the process that can be fairly sampled. However, as already noticed for the previous example, by increasing $N$ of $5$ orders of magnitude we only get to double the \deletetext{length} \newtext{duration} of the process that is well sampled. This exponential relation, i.e., $N \propto e^n$, is natural in sampling large deviations of additive observables.

In Fig.\ \ref{fig:scgf3} we plot the estimated SCGFs given by \eqref{eq:SCGFEst} for different values of $n$ at fixed $N=10^5$, and compare them with the true SCGF $\Psi(k)$ and vertical lines corresponding to critical values of $k_c$. As expected, even in this case estimates are good up to a certain value of $k$ after which linearisation effects take place. Furthermore, for values of $n \gtrsim 25$ we can predict with great accuracy the values of $k_c$. Estimates reveal that the larger the $n$, the worse is the convergence to IFT (at $k=-1$). 

\begin{figure}
    \centering
    \includegraphics[width=0.5\linewidth]{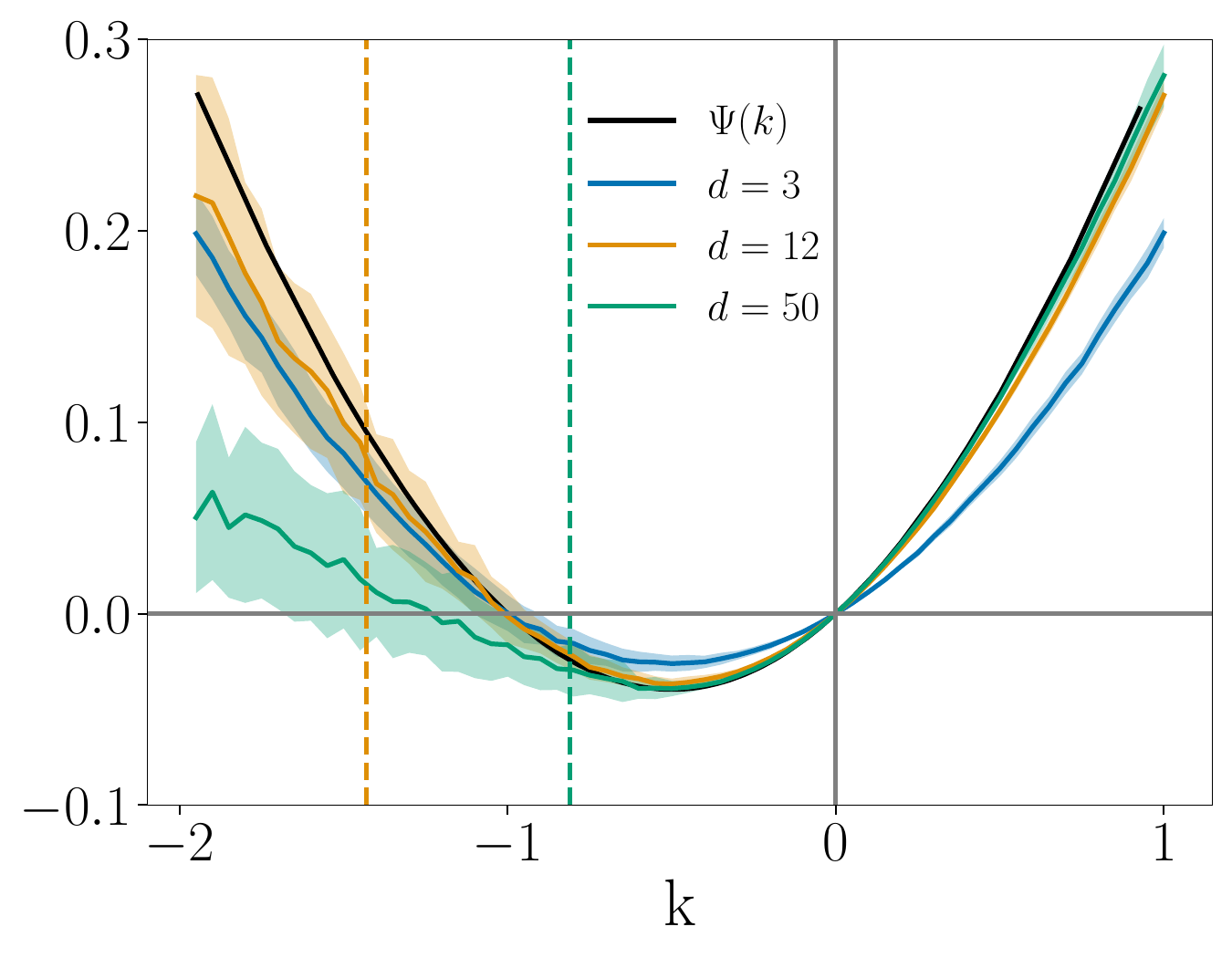}
    \caption{Estimated SCGFs (coloured solid curves) given by $\hat{\Psi}_N(k)$ in \eqref{eq:SCGFEstCorr} for different values of $d$ with $N=10^3$ copies (the shaded region corresponds to one standard deviation computed over $100$ samples) compared with the true SCGF $\Psi(k)$ (black solid curve) \newtext{and with values of $k_c$ (coloured vertical dashed lines) with varying $d$ (in particular, for $d=\xi$ we have $k_c=k_{c,y}$ in \eqref{eq:kcCorrBetter}}).}
    \label{fig:scgf_trick_3}
\end{figure}

Once again, we show that we can get rid of the bias and the dependence on $n$ by exploiting the fact that the Markov chain has a finite correlation \deletetext{length} \newtext{time} $\xi$. We estimate this by calculating the lag-$d$ correlation \deletetext{length} \newtext{time} of the chain~\cite{Levin2017}, i.e.,
\begin{equation}
\label{eq:RealThreeStateLagdCorrLength}
l(d) = \mathbb{E}\left[ Z_n Z_{n+d} \right] = \sum_{z,z'} \mu_0^f(z) z z' \,\Pi^d(z,z') \ ,
\end{equation}
and by taking $\xi=d$ to be the first value for which the correlation is $0$. [Notice that the second equality of \eqref{eq:RealThreeStateLagdCorrLength} is only valid for $n$ larger than the relaxation time of the chain. \newtext{As previously mentioned, we initialise our Markov chain in its stationary distribution and therefore we can safely use \eqref{eq:RealThreeStateLagdCorrLength}.}] We estimate $l(d)$ for our simple model and see that for the particular case of $p_{-1}$, $p_0$, and $p_1$ chosen, $\xi \approx \newtext{10\sim}12$. We then consider \eqref{eq:MomentActFunctCorr} and the estimator \eqref{eq:EstMomentCorr} with $Y_\ell$ i.i.d.\ variables given by \eqref{eq:YellClustered}. By looking at Fig.\ \ref{fig:scgf3} we know that to have good convergence at $k=-1$ for $n=10$ we only need $N \approx 10^\newtext{3}$. We plot in Fig.\ \ref{fig:scgf_trick_3} the new estimate of the SCGF given by \eqref{eq:SCGFEstCorr} \newtext{for different values of $\xi$.}\deletetext{ which, once again, confirms the great improvement we can achieve by exploiting the finite correlation length of the process.} \newtext{For $d<\xi$ the chain is too short and it has not decorrelated over time yet. As a consequence, the blue line lies far from the true (black) one---although well sampled and therefore in good agreement with IFT---and cannot be considered an estimate of the real SCGF. On the other hand, if $d>\xi$, by keeping $N$ the smallest possible for $d=\xi$ one may incur into downsampling and linearisation effects as shown in the plot. Furthermore, the plotted dashed lines correspond to the critical values $k_c$ at fixed $d$ and $N$. Evidently, \eqref{eq:kcCorrBetter} is now satisfied with $k_{c,y}$ given by $k_c$ for $d=\xi$ and fixed $N$.}

\section{Conclusions}
\label{sec:concl}

In this paper, by focusing on discrete-time Markov chains, we have discussed how the IFT estimator behaves when applied to equilibrium and nonequilibrium systems. We have seen that for the former systems no matter the sampling size, the estimator always converges to $1$, whereas for nonequilibrium systems the estimator may be strongly biased due to undersampling of rare events which lead to linearisation effects in the SCGF estimator. Sampling rare events is indeed pivotal when it comes to statistically estimate IFT for nonequilibrium systems as these lasts typically have an entropy production greater than zero, which tilts the IFT estimator towards values smaller than one if a careful sampling of all rare events is not carried out. As thoroughly discussed in the paper, this and similar sampling problems are already well known in the literature, and it is also clear that regular sampling carries an exponential complexity, viz.\ the number of samples required to sample rare events scale at least exponentially with the time \deletetext{length} \newtext{duration} of the system. Nevertheless, by applying arguments developed in \cite{Duffy2005,Rohwer2015} we have discussed in this paper how to determine the convergence window of the IFT estimator as a function of both the \deletetext{length} \newtext{duration} of the investigated process and the sample size and, in particular, how to \textit{exponentially} improve the convergence of the estimator for Markov chains that have a finite correlation \deletetext{length} \newtext{time}. \newtext{Interestingly, although we have only focused on the action functional (or entropy production rate in our case) the results obtained are general for time-extensive observables, e.g., stochastic heat and work just to mention relevant observables for stochastic thermodynamics. It is indeed always the case that time extensiveness plays against sampling accuracy: the longer the trajectory to sample, the harder it will be to sample rare events.}


Recently, much progress has been done on methods to sample rare events of additive observables like the action functional we have treated in this work. These methods (not discussed here) make use of importance sampling~\cite{Kundu2011,Ray2018,Coghi2022}, particle filters~\cite{DelMoral2004}, such as cloning~\cite{Giardina2006,Lecomte2007} and splitting algorithms~\cite{Cerou2007,Dean2009}, and reinforcement~\cite{Rose2021} and machine learning~\cite{Yan2022}. Although less efficient, regular sampling is still of strong interest. This is because---mainly when it comes to experiment---it is more straightforward to implement. \newtext{However, tools to analyse the convergence properties of these ordinary estimators---including the ones discussed in this paper---are still very much based on techniques that require knowledge on the system, e.g., tilted transition matrices, that is hard to assess experimentally.} An interesting research avenue in the authors' opinion would indeed be to investigate and `adapt' the aforementioned advanced numerical methods \newtext{along with those discussed in this manuscript} to experimental settings.

In addition to what we have discussed so far, effects of limited sampling in nonequilibrium thermodynamics have also been recently investigated in \cite{buffoni2022spontaneous}, in reference with the way the Landauer bound \cite{BerutNature2012,DasPRE2014,DagoPRL2021} emerges as a consequence of a non-convergence of the Jarzynski equality estimator for an Ising model with a large number of spins. 
%
%
%
In applications, the non-convergence of the Jarzynski equality estimator, the IFT estimator, or any other statistical estimator that deals with sums of exponentials, is usually `fixed' by adding `ad-hoc' terms that make the estimator converge to the desired value, see~\cite{JunierPRL2009,AlemanyNatPhys2012} for biophysics' related works. Such `ad-hoc' corrections arise whenever rare (but fundamental) trajectories of the system dynamics are not sampled and hence do not contribute to the empirical average in \eqref{eq:CFTEstimator}---something that in certain scenarios could also be related to absolute irreversibility~\cite{MurashitaPREnonequilibrium}. We therefore argue that these corrections to the statistical estimator could more easily be understood---and, perhaps, avoided---by applying the methods and results discussed in this paper.




%
%


\section*{Acknowledgments}

The authors acknowledge The Blanceflor Foundation for financial support through the project ``Large Deviations approach to Landauer’s Principle (LanDev)'' and the MISTI Global Seed Funds MIT-FVG collaboration grant ``Non-Equilibrium Thermodynamics of Dissipative Quantum Systems''. We also thank the International Centre for Theoretical Physics (ICTP) in Trieste (Italy) for hospitality during the completion of this work. FC gratefully acknowledges Centre National de la Recherche Scientifique (CNRS) and the Laboratoire de Physique at the ENS Lyon (LPENSL) for hospitality during the writing stage of the manuscript. The authors also warmly thank Hugo Touchette for providing useful comments and suggestions on a first draft of the work.

\bibliographystyle{unsrt}
\bibliography{bibliography}


\appendix

\section{Derivation of the end-point solution $\bar{m}_c(n,N)$ in Eq.~(\ref{eq:ThreeStateEndPoint})}
\label{sec:appmc}

Following \eqref{eq:EndPointDef} and \eqref{eq:CumulativeDistribution} we get
\begin{equation}
    \label{eq:ThreeStateCumulative}
    F_n(\bar{m}_c) = \sum_{m=0}^{\bar{m}_c} \mathbb{P}_n(M_c=m) = \sum_{m=0}^{\bar{m}_c} {n \choose m} p^{m} (1-p)^{n-m} \ .
\end{equation}
In the $n \rightarrow \infty$ (in practice, $n$ big) we can approximate the binomial distribution of \eqref{eq:ThreeStateBinDistr} with a normal distribution with mean $np$ and variance $np(1-p)$. Within this approximation, the cumulative distribution function in \eqref{eq:ThreeStateCumulative} reads
\begin{equation}
    \label{eq:ThreeStateCumulativeNormal}
    F_n(\bar{m}_c) \approx \int_{0}^{\bar{m}_c} dm \,  \frac{1}{\sqrt{2 \pi n p (1-p)}} e^{- \frac{(m - np)^2}{2 n p (1-p)}} \ ,
\end{equation}
where we also put $0$ (instead of $-\infty$) as the left extremum in the integral with the assumption that $p \sim O(1)$. [Should $p$ scale inversely with respect to $n$ the whole argument here is no longer valid.]

For simplicity, we stick with the condition $p<1/2$. [All the argument can be adjusted for the case $p>1/2$.] In such a case, the rare events we are interested into and that define the end point $\bar{m}_c$, are those for $\bar{m}_c \gg np$. We rewrite \eqref{eq:ThreeStateCumulativeNormal} introducing the error function as follows:
\begin{equation}
    \label{eq:ThreeStateCumulativeErf}
    F_n(\bar{m}_c) = \frac{1}{2} \left[ 1 + \text{erf} \left( \frac{\bar{m}_c - np}{\sqrt{2 np (1-p)}} \right) \right] \ .
\end{equation}
By reverting \eqref{eq:ThreeStateCumulativeErf} we get
\begin{equation}
    \label{eq:ThreeStateBoundaryPoint}
    \begin{split}
        \bar{m}_c(n,N) &= \sqrt{2n p (1-p)} \text{erf}^{-1} \left( 2 F_n(\bar{m}_c)-1 \right) + np \\
        &= \sqrt{2n p (1-p)} \text{erfc}^{-1} \left( 2-2 F_n(\bar{m}_c) \right) + np \ , 
    \end{split}
\end{equation}
where, at last, we have introduced the complementary error function ($\text{erfc}(z) = 1 - \text{erf}(z)$). Now, as we are looking at $m_c$ rather than $s$, we would like $F_n(\bar{m}_c) \rightarrow 1 \approx 1 - \tau/N$. By replacing this condition in \eqref{eq:ThreeStateBoundaryPoint} we get
\begin{equation}
    \label{eq:ThreeStateBoundaryPointExpl}
        \bar{m}_c(n,N) =  \sqrt{2n p (1-p)} \text{erfc}^{-1} \left( \frac{2 \tau}{N} \right) + np \ , 
\end{equation}

We now make use of the well-known asymptotic formula for the inverse complementary error function, that is
\begin{equation}
    \label{eq:FormulaAsympErfc}
    \text{erfc}^{-1}(x) = u^{-1/2} + o(u^{-1/2}) \ ,
\end{equation}
with 
\begin{equation}
    u = - \frac{2}{\ln (-\pi x^2 \ln(x))} \ ,
\end{equation}
for $x \rightarrow 0$. Applied in our case, we get
\begin{equation}
    \label{eq:ThreeStateAsymp}
    \text{erfc}^{-1} \left(\frac{2 \tau}{N} \right) \approx \sqrt{-\frac{1}{2} \left( \ln \pi - 2 \ln \left( \frac{N}{2 \tau} \right) + \ln \ln \left( \frac{N}{2 \tau} \right) \right)} \approx \sqrt{\ln \left( \frac{N}{2 \tau} \right)} \ .
\end{equation}

Eventually, by replacing \eqref{eq:ThreeStateAsymp} in \eqref{eq:ThreeStateBoundaryPointExpl}, we obtain $\bar{m}_c(n,N)$ to equate to \eqref{eq:ThreeStateSaddle} to get \eqref{eq:ThreeStateKcToStudy}.

\end{document}